\documentclass[10pt,twocolumn,letterpaper]{article}

\usepackage[pagenumbers]{cvpr} 

\usepackage{graphicx}
\usepackage{subcaption}
\usepackage{amsmath,bm}
\usepackage{amssymb}
\usepackage{times}
\usepackage{booktabs}
\usepackage{diagbox}
\usepackage{multirow}
\usepackage{caption}
\usepackage{float} 
\usepackage{cuted}
\usepackage{sidecap}

\definecolor{cvprblue}{rgb}{0.21,0.49,0.74}
\usepackage[pagebackref,breaklinks,colorlinks,allcolors=cvprblue]{hyperref}

\def\paperID{15188} 
\def\confName{CVPR}
\def\confYear{2025}
\title{GraphicsDreamer: Image to 3D Generation with Physical Consistency}

\author{Pei Chen$^{1}\footnotemark[1]$,
Fudong Wang$^{1}\footnotemark[1]$,
Yixuan Tong$^{1,2}$,
Jingdong Chen$^{1}$,
Ming Yang$^{1}$
Minghui Yang$^{1}$
\vspace{0.2cm}
\\
{\normalsize $^{1}$ Ant Group \quad $^{2}$ Fudan University}
\vspace{0.2cm}
\\
}

\begin{document}
\maketitle

\footnotetext[1]{Equal contribution.}

\begin{strip}
    \centering
    \centering
    {\includegraphics[width=0.95\linewidth]{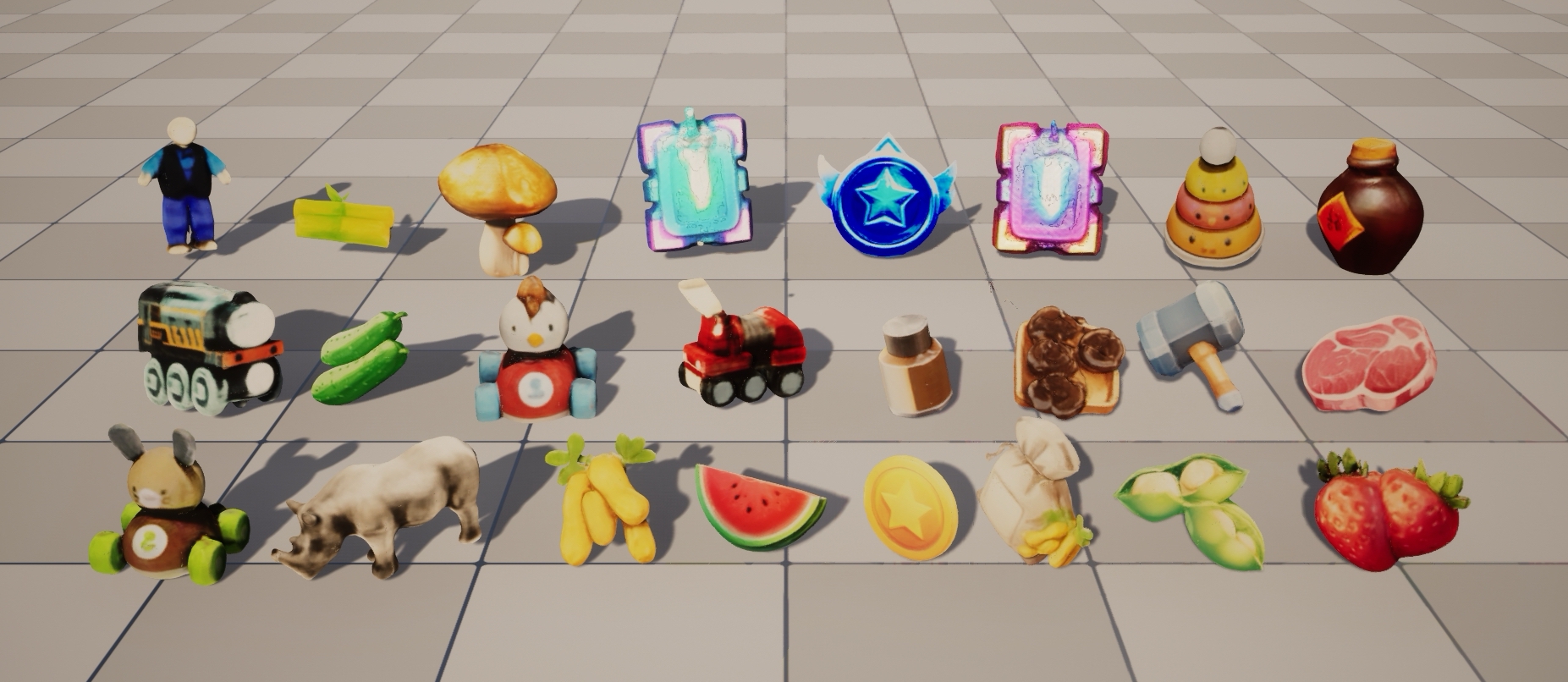}} 
    \captionof{figure}{\textit{GraphicsDreamer} utilizes a two-stage generation approach, integrating PBR lighting conditions into both the multi-view synthesis and reconstruction processes. The 3D models produced by \textit{GraphicsDreamer} possess clear geometry, clean topology, and complete PBR maps, allowing them to be directly imported and manipulated within graphic engines.}
    \label{fig:teaser}
\end{strip}

\begin{abstract}

Recently, the surge of efficient and automated 3D AI-generated content (AIGC) methods has increasingly illuminated the path of transforming human imagination into complex 3D structures. However, the automated generation of 3D content is still significantly lags in industrial application. This gap exists because 3D modeling demands high-quality assets with sharp geometry, exquisite topology, and physically based rendering (PBR), among other criteria. To narrow the disparity between generated results and artists' expectations, we introduce \textbf{GraphicsDreamer}, a method for creating highly usable 3D meshes from single images. To better capture the geometry and material details, we integrate the PBR lighting equation into our cross-domain diffusion model, concurrently predicting multi-view color, normal, depth images, and PBR materials. In the geometry fusion stage, we continue to enforce the PBR constraints, ensuring that the generated 3D objects possess reliable texture details, supporting realistic relighting. Furthermore, our method incorporates topology optimization and fast UV unwrapping capabilities, allowing the 3D products to be seamlessly imported into graphics engines. Extensive experiments demonstrate that our model can produce high quality 3D assets in a reasonable time cost compared to previous methods.
\end{abstract}
\section{Introduction}
\label{sec:intro}

Traditional 3D modeling processes heavily relies on manual labor, which significantly hinders the application of 3D content in Internet and gaming industries. Excitingly, with the evolution of generative AI, many works \cite{liu2023syncdreamer, jun2023shap, qian2023magic123, melas2023realfusion, dou2023tore, liu2023zero,Long_2021_ICCV, nichol2022point} have emerged that attempt to quickly produce 3D models from text or images, showcasing the potential to free human efforts from this complex and labor-intensive task of 3D modeling. However, achieving this goal is challenging as the 3D models qualified for direct integration into rendering engines must meet high graphics standards in terms of geometry and materials.

In terms of \textbf{geometry}, a desired 3D model should have sharp, clear edges and a concise topology to be efficiently rendered and edited in graphics engines. Models with blurry geometric features, messy topology, and excessively high polygon counts are difficult for human artists to further edit, making them practically useless. In terms of \textbf{materials}, we certainly hope that each 3D character, object can display reasonable light and shadow effects under different lighting conditions, which can be achieved with PBR materials.

As the field of image generation sees more impressive works emerge, an intuitive method is to distill prior knowledge from image diffusion models and use Score Distillation Sampling (SDS) \cite{poole2022dreamfusion} to generate 3D models from text or images, such as DreamFusion \cite{poole2022dreamfusion}, Fantasia3D \cite{chen2023fantasia3d}, Magic3D \cite{lin2023magic3d}, and ProlificDreamer \cite{wang2023prolificdreamer}. While these pioneers have demonstrated the feasibility of this approach, they suffer from two limitations. The first is slow inference speeds because the per-shape optimization involves tens of thousands of iterations, often requiring tens of minutes or even hours. The second issue, known as the 'Janus problem' \cite{poole2022dreamfusion,wang2023score}, arises because the model, in each iteration, strives to align the current view with the input image, often resulting in objects with multi-faces that severely impact the model robustness.

Another strand of research directly uses 3D data as training data, training a 3D generative model from scratch to perform end-to-end generation of structures like voxels \cite{wu2016learning,henzler2019escaping}, point clouds \cite{nichol2022point, zeng2022lion, luo2021diffusion, zhou20213d}, meshes \cite{liu2023meshdiffusion,gao2022get3d}, and neural fields \cite{wang2023rodin, cheng2023sdfusion,gupta20233dgen,erkocc2023hyperdiffusion}. However, due to the limitation of publicly accessible 3D data, these models have weak generalization capabilities and low fidelity, and often generating 3D structures not present in the input images.

To address the issue of consistency, subsequent methods directly generate multi-view consistent 2D images and reconstruct 3D geometry from them, such as SyncDreamer \cite{liu2023syncdreamer} and MVDream \cite{shi2023mvdream}. Fantasia3D \cite{chen2023fantasia3d} attempts to decouple geometry and textures, as well as PBR materials from ambient lighting. However, since in 2D images, the object’s geometry, texture, and lighting information are naturally intertwined, relying solely on color images to optimize normal maps leads to unstable optimizations due to differing data distributions, resulting in significant detail loss. Wonder3D \cite{long2023wonder3d} introduces normal maps in the training data and uses a cross-domain diffusion model to generate multi-view consistent color and normal images for 3D reconstruction, enhancing the detail of the 3D results but lacking a comprehensive understanding of 3D structures. RichDreamer \cite{qiu2024richdreamer} utilizes a Normal-Depth diffusion model to similarly control and generate across multiple domains and employs an albedo diffusion model to mitigate the interference of mixed illumination, yet it has not evolved to the generation of complete PBR material maps.

\begin{figure*}[tp!]
	\centering
	{\includegraphics[width=1.0\linewidth]{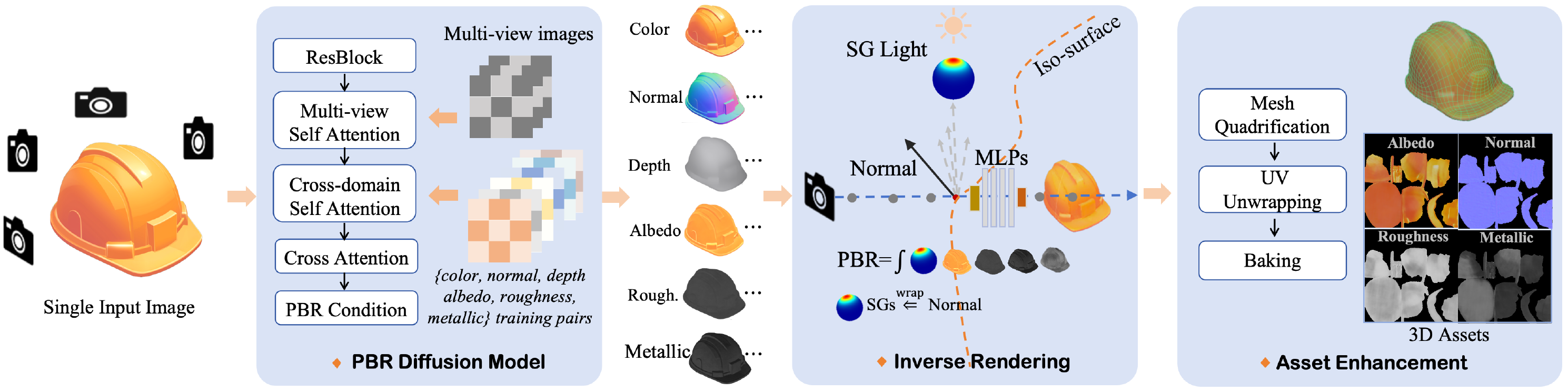}}
    \vspace{-3mm}
    \captionof{figure}{Our method consists of three phases. Given a single input image, we train a diffusion model (Sec.~\ref{sec:sprseview_gen}) to generate multi view images (6 views), including RGB color for the overall appearance, normal and depth as geometric information, intrinsic materials for texture details, conditioned by a PBR approch (Sec.~\ref{sec:pbr_render}). The generated images are then integrated into an inverse rendering reconstruction (Sec.~\ref{sec:explit_iso}) also in conjunction with the PBR process to guarantee the consistency with the diffusion model. At last, our method will product appealing 3D assets with artistically optimized topology and UV textures (Sec.~\ref{sec:assts_3d}).}
    \vspace{-5mm}
	\label{fig:pipeline}
\end{figure*}

Our method also adopts a two-stage scheme, initially generating multi-view images, followed by reconstruction, and incorporates the PBR lighting condition into each stage. Recognizing that depth information encapsulates a comprehensive understanding of the scene’s overall geometric structure, normal maps represents the surface details of 3D objects, and PBR materials provide a rich description of the object’s surface textures, we expand the first stage into a \textit{\textbf{PBR diffusion model}}. This model predicts the joint distribution of six domains, including colors, normals, depths, and PBR components (albedo, roughness, metallic), as shown in Fig.~\ref{fig:nvs}. It uses a cross-domain diffusion model similar to Wonder3D \cite{long2023wonder3d} as its basic structure and integrates the PBR lighting conditions. To aggregate generated multi view images and intrinsic materials into a complete 3D object as a surface mesh, we further introduce a deep learning-based \textit{\textbf{inverse rendering}} approach consists of a mixed representation of implicit and explicit surface, also along with a PBR-constrained intrinsic material reconstruction. With the support of the PBR lighting module, our model can easily distinguish highlights and shadows from the actual surface textures of 3D objects, see Fig.~\ref{fig:ablation}.

 Furthermore, to align with the computer graphics pipeline, we have automated the topology optimization and UV unwrapping of 3D objects, allowing them to be directly imported into rendering engines such as Blender \cite{Blender}, Unreal Engine, and Unity. Extensive testing on the Google Scanned Object (GSO) \cite{downs2022google} dataset has shown that, compared to other baseline methods, GraphicsDreamer can produce high-quality 3D meshes with photorealistic textures.

The core contributions of this paper are as follows:

\begin{itemize}

\item GraphicsDreamer integrates both geometry and materials generation into the multi-view diffusion model, which is further enhanced by a PBR condition with environment lighting approximated as spherical Gaussian. This enhanced model provides a wealth of usable information for the subsequent reconstruction phase.


\item GraphicsDreamer proposes a deep learning-based inverse rendering approach consists of a mixed surface representation and a PBR-constrained intrinsic material enhancement. The resulting complete 3D surface mesh features smooth geometry and distinct textures since it leverages both the generalization capability of the diffusion model and the refinement ability of inverse rendering.

\item GraphicsDreamer incorporates capabilities for topology optimization and UV unwrapping, which are often neglected in academic 3D generation methods. Experiments have shown that in terms of geometric and texture details, GraphicsDreamer is at a leading level. Moreover, the complete PBR material maps and clean topology allow the generated 3D models to be directly imported into graphics engines for immediate use.

\end{itemize}

\section{Related Work}

\subsection{2D Diffusion for 3D Generation}

Recent advancements has demonstrated that utilizing CLIP model \cite{radford2021learning,jain2022zero,xu2023dream3d,mohammad2022clip} or a 2D diffusion model \cite{saharia2022photorealistic,rombach2022high}, researchers can directly generate 3D objects from user prompts. The pioneer work DreamFusion \cite{poole2022dreamfusion} leverages score distillation sampling (SDS) to extract prior knowledge from a 2D diffusion model, iteratively optimizing a neural radiance field (NeRF) \cite{mildenhall2020nerf} and achieve zero-shot text-to-3D generation. Concurrently, SJC \cite{wang2023score} utilizes score Jacobian chaining to achieve a similar goal. Building on this foundation, Magic3D \cite{lin2023magic3d} employs an improved multi-resolution SDS to enhances the precision of 3D generation. Fantasia3D \cite{chen2023fantasia3d} integrates DMTet and SDS, separating textures from geometry, aiming to improve texture quality.

However, due to the lack of multi-view constraints, these methods often produce objects with multi faces. Additionally, the hour-level generation time required for per-shape optimization is usually difficult to accept. Other methods, such as One-2-3-45 \cite{hugging2023one2345}, Magic123 \cite{qian2023magic123}, and Make-it-3d \cite{tang2023makeit3d}, directly generate 3D geometries from image conditions, though they significantly speed up generation, the quality is lower, lacking in geometric and texture details.

\subsection{3D Generative Models}

This type of generative model is trained directly on 3D data, learning to capture the distribution of 3D data, and have achieved convincing results. The forms of 3D representations can generally be classified into voxels \cite{wu2016learning,henzler2019escaping}, point clouds \cite{nichol2022point, zeng2022lion, luo2021diffusion}, meshes \cite{liu2023meshdiffusion,gao2022get3d}, and neural fields \cite{wang2023rodin,cheng2023sdfusion, muller2023diffrf, zhang20233dshape2vecset, gupta20233dgen}. However, limited by the scale of available 3D training data, such models are often restricted to generating objects within certain specific categories and frequently fabricate elements not present in the input images. In contrast, our method uses a 2D representation across $6$ domains to model 3D objects, ensures better zero-shot capabilities and faithfully reproduces the 3D geometry according to the input images.

\begin{figure*}[tp!]
	\centering
	{\includegraphics[width=1.0\linewidth]{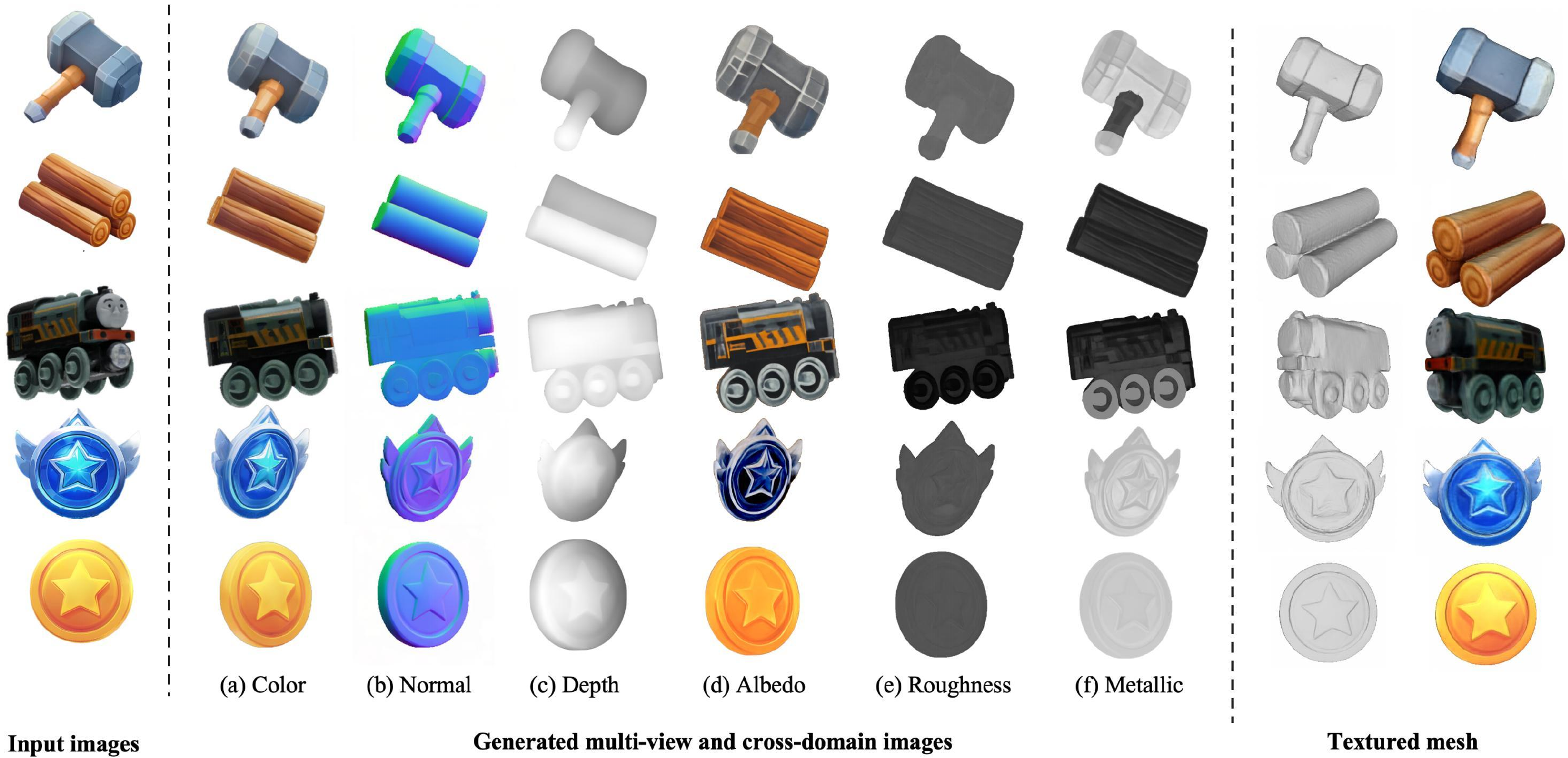}} 
    \captionof{figure}{The novel-view synthesis results by the \textbf{\textit{GraphicsDreamer}} on objects with different materials. As can be seen, the model effectively separates the inherent colors of objects under various lighting conditions, as observed in the textures on the \textbf{\textit{train}} and \textbf{\textit{coins}} in the \textbf{albedo} column. It also identifies metallic materials, such as the \textbf{\textit{hammer}} and \textbf{\textit{train wheels}} in the \textbf{metallic} column. This capability to identify such characteristics is essential for precisely conveying the material of objects and for successful relighting.}
    \vspace{-5mm}
	\label{fig:nvs}
\end{figure*}

\vspace{-2mm}
\subsection{Multi-view Diffusion Models}

Zero-1-to-3 \cite{liu2023zero} enhances the 2D diffusion by fine-tuning the Stable Diffusion model \cite{rombach2022high}, enabling it to perform novel-view synthesis from specified views. More recent developments have seen significant improvements in the consistency of multi-view image generation through multi-view diffusion \cite{shi2023mvdream,liu2023syncdreamer,ye2023consistent,weng2023consistent123,shi2023zero123++,tseng2023consistent,zhao2023efficientdreamer,szymanowicz2023viewset}. A prominent project in this series is MVDream \cite{shi2023mvdream}, which fine-tunes a pretrained diffusion model using multi-view images rendered from 3D objects in the Objaverse dataset \cite{deitke2023objaverse}. However, because they rely solely on RGB images, often encounter texture ambiguity during geometric reconstruction.

Subsequent work, Wonder3D \cite{long2023wonder3d} incorporates normal images into its training data, utilizing an RGB-Normal diffusion model to enhance the generated geometric details. RichDreamer’s \cite{qiu2024richdreamer} Normal-Depth diffusion model enables it to generate rich geometric details. In contrast, our model constructs 3D objects as a joint distribution of images in six domains, comprehensively representing both geometry and materials. By integrating the PBR conditions into both the multi-view image synthesis stage and geometry fusion stage, our approach significantly boosts the generation quality, fully adhering to the computer graphics pipeline.

\subsection{Materials and Light Estimation}
Known as intrinsic image decomposition~\cite{intri78, Intrinsic2022} or inverse rendering~\cite{inverse2003}, it's a challenging task to estimate the geometry, intrinsic materials and lighting of observed objects with single or multi view images based on physics principles~\cite{pbr_3ed}. The ill-posed nature of this problem demands that multiple values (normal, albedo, roughness, metallic, specular, illumination, etc.) per pixel should be calculated with only several correspondent RGB values observed cross views. The early literature~\cite{intri78,intri1971,inverse2003} can solve some simplified situations with specific priors, and some recent works aim to enhance the captured images using more sophisticated capture systems~\cite{invehuman2019} or controllable lighting environment~\cite{inve_flash2018}. Inspired by the development of modern deep learning, many works~\cite{intri_ordinary23, Intrinsic2022, invertransformer2022,intranything2024,mgnet2022} aim to learn the decompositions directly from images with neural network models, which are trained upon elaborately constructed datasets that consist of indoor scenes~\cite{intridata2014,intridata2017, inverdata2020,openroom2021} or objects~\cite{stanfordorb2023,matathome2021}. Most of them focus on learning intrinsic priors from single image and then estimate the lighting with an optimization method if needed. 

Moreover, the decomposition of intrinsics and lighting with implicit geometry representations e.g., neural radiance filed, has garnered significant attention in recent years to take advantages of the more flexible differentiability. These works~\cite{invnerf++2023, physg2021, idr2020, nvdifmc2022,dvr2020,neus2021} inspired by Nerf~\cite{nerf2020} reformulate the imaging process using intrinsic materials and lighting accompanied with implicit geometry representation like signed distance function (SDF). Some latest studies~\cite{GS_avator_relht2024, Rlight3DGS2023} go further utilize the 3D gaussian splatting-based representation and then optimize the intrinsic materials. For the lighting representation, spherical Gaussian~\cite{SG_close2018, physg2021, SGlight2009, 2024iid} and its variant~\cite{SVBRDF2020} are primary representations that can recover higher-frequency reflection compared with spherical harmonic lighting~\cite{SH_gl2019}.
\vspace{-3mm}
\section{Method}

As illustrated  in Fig.~\ref{fig:pipeline}, GraphicsDreamer consists of three phases. Given a single input image of the desired object, we first build a diffusion model to generate multi view images (6 views), including RGB color for the overall appearance, normal and depth as geometric information, intrinsic materials for texture details, which are simultaneously controlled by a PBR condition as demonstrated in Sec.~\ref{sec:sprseview_gen}. The generated images are then treated as pseudo ground truth for refining a deep learning-based inverse rendering reconstruction, which is also conditioned by PBR to keep consistent with the generative model. It results in a complete 3D object as a surface mesh with smooth geometry and distinct textures, as described in Sec.~\ref{sec:explit_iso} and Sec.~\ref{sec:pbr_render}. At last, our method will product appealing 3D assets with artistically optimized topology and UV textures, which are essential in modern CG workflow, as shown in Sec.~\ref{sec:assts_3d}. 

\subsection{PBR Diffusion Model}\label{sec:sprseview_gen}
\noindent
\textbf{The Distribution of 3D Assets.} Previous work, such as Wonder3D \cite{long2023wonder3d} and RichDreamer \cite{qiu2024richdreamer}, selected color images along with corresponding normal or depth images as learning targets for their 2D diffusion models. While this representation can model 3D geometry, it falls short in adequately decoupling intrinsic textures and lighting effects, which affects the quality of relighting results. 

Therefore, to better decouple geometry, materials, and lighting, and to make the appearance of the 3D content more realistic, we propose modeling 3D assets as a joint distribution of color images and corresponding normal, depth images, as well as material component maps. Specifically, the distribution of 3D assets denoted as $p_a(\mathbf{z})$ is defined as
\begin{equation}
\label{eqn:joint_distribution}
p_a(\mathbf{z})=p_{pbr}\left(c^{1:K}, n^{1:K}, d^{1:K}, a^{1:K}, r^{1:K}, m^{1:K} | y \right)
\end{equation}
Here, $p_{pbr}$ refers to the distribution of the 3D asset’s multi-view images, which are colors $c^{1:K}$, normals $n^{1:K}$, depths $d^{1:K}$, albedos $a^{1:K}$, roughnesses $r^{1:K}$, and metallics $m^{1:K}$ under the condition of input viewpoint $y$. $K$ denotes the number of camera views, which is set to 6 in our experiments. Thus, our objective is to train a model $f$ capable of predicting multi-view images across these six domains, given a single input view $y$ and a fixed set of camera configurations $\boldsymbol{\pi}_{1:K}$:
\begin{equation}
\label{eqn: model_func_0}
(c^{1:K}, n^{1:K}, d^{1:K}, a^{1:K}, r^{1:K}, m^{1:K})=f(y, \boldsymbol{\pi}_{1:K})
\end{equation}

\noindent
\textbf{Multi-view Diffusion for Geometry and Materials.} 
Similar to MVDream \cite{shi2023mvdream} and Wonder3D \cite{long2023wonder3d}, we employ a multi-view self-attention mechanism. Previous methods have proven that by connecting keys and values of different views within the attention layer for information sharing, the model’s capacity for 3D global perception is enhanced and results in significant improvement in the consistency of multi-view generation. 

Then, we refine Wonder3D’s \cite{long2023wonder3d} cross-domain attention network to accommodate more domains. We utilize a domain switcher $s\in\{0, 1\}$ to label different domains. The switcher is then encoded and concatenated with input images and camera parameters before feeding it into the UNet of the diffusion model for training. Experiments show that this design preserves the prior knowledge of the pretrained model and supports fast convergence and robust generalization, even with the expansion to six domains. 

The key challenge, however, is to ensure that the six domain images generated from a single view are geometrically consistent. To tackle this, we first recognize that color images most completely reflect an object’s appearance, while normal, depth, albedo, roughness, and metallic can be considered the most fundamental, indivisible atomic properties that contribute to the final appearance of an object, that is, the color image. Consequently, we treat the color image domain as the primary domain, using Query $Q_c$ from the color domain to calculate cross-attention $\alpha_i$ separately with Keys $K_i$ from the other five domains, which is
\begin{equation}
\label{eqn: cross_attention}
\alpha_i = \text{softmax}\left({Q_c \cdot K_i}/{\sqrt{d_k}}\right),
\end{equation}
where $i$ represents the other five domains, and $d_k$ is the dimension of the $K_i$ vectors. Following this, we employ a single-layer MLP to perform a weighted fusion of the cross-attention weights. 

Then, as a process related to lighting and PBR, we use the rendering equation detailed in Eqn.~\ref{eq:pbr_eq0} to supervise whether the color images are accurately represented by the four component maps: normal, albedo, roughness, and metallic. This supervision is crucial for achieving photorealistic rendering and enhances the interpretability of the model. For the specific mathematical process, see Sec.~\ref{sec:pbr_render}. 

\begin{figure*}[tp!]
	\centering
	{\includegraphics[width=1.0\linewidth]{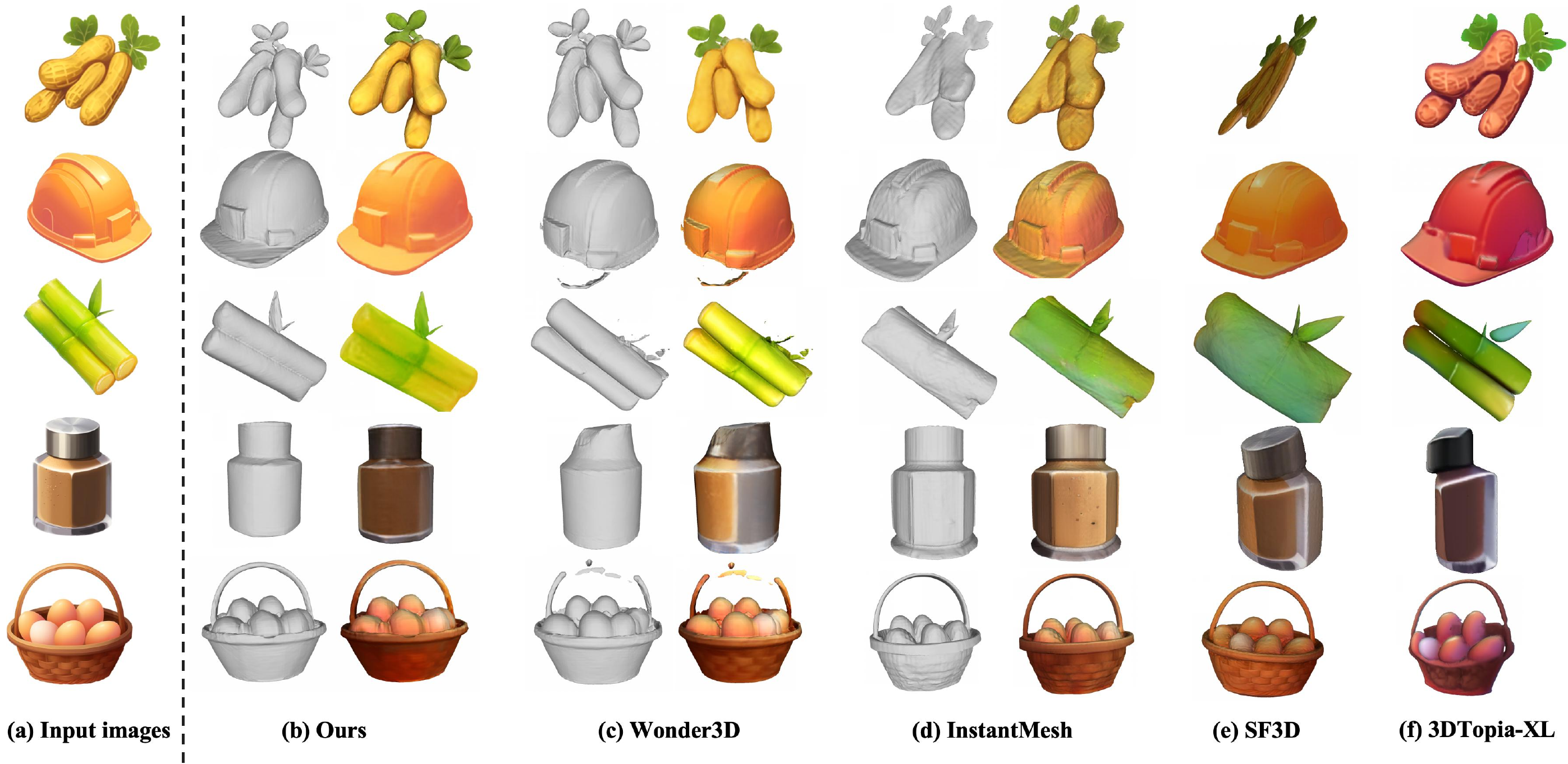}} 
    \captionof{figure}{The qualitative comparisons with baseline methods on single view reconstruction. Due to the line width limitation, we do not show the untextured meshes for SF3D and 3DTopia-XL.}
    \vspace{-5mm}
	\label{fig:recon}
\end{figure*}

\subsection{Mixed Surface Representation}\label{sec:explit_iso}
With the above generated sparse-view images and materials of object, we need to recover the geometry based on which the physically-based rendering and lighting procedure is performed.
Has been proved its efficiency in current studies~\cite{nerf2020, neus2021} that implicit representation, {\eg} signed distance function (SDF), can achieve better differentiability and stability, it is incompatible with the reflection occurring on object's surface incorporated in PBR function and thus can not be used directly. In this section, we introduce a mixture representation consists of implicit SDF and explicit surface such that both the differentiability, stability and compatibility can be achieved simultaneously.

\noindent
\textbf{Implicit SDF Initilization.} We initially adopt the implicit SDF representation introduced in NeuS~\cite{neus2021} that can convergent to the zero-level iso-surface $\mathcal{S}$ faster. Given a ray $\mathbf{r}\triangleq \mathbf{r}_o + t\cdot \mathbf{r}_d$, where $\mathbf{r}_o, \mathbf{r}_d\in \mathbb{R}^3$ are the ordinary and destination respectively, $t\in \mathbb{R}^+$ is the depth of the current sample distance, we define the SDF $f\in \mathbb{R}$ of a sampled point $\mathbf{x}\triangleq\mathbf{r}_o + t_x\cdot \mathbf{r}_d$ on the ray $\mathbf{r}$ as $f(\mathbf{x})\triangleq f_m(\theta, \mathbf{x})$, where $f_m(\theta, \cdot)$ consists of MLPs w.r.t. the neural parameters $\theta$. The zero-level isosurface $\mathcal{S}$ of the desired object is $f_m(\theta, \mathbf{x})=0 \Leftrightarrow \mathbf{x}\in \mathcal{S}$. 

\begin{figure}
	\centering
    {\includegraphics[width=1.0\linewidth]{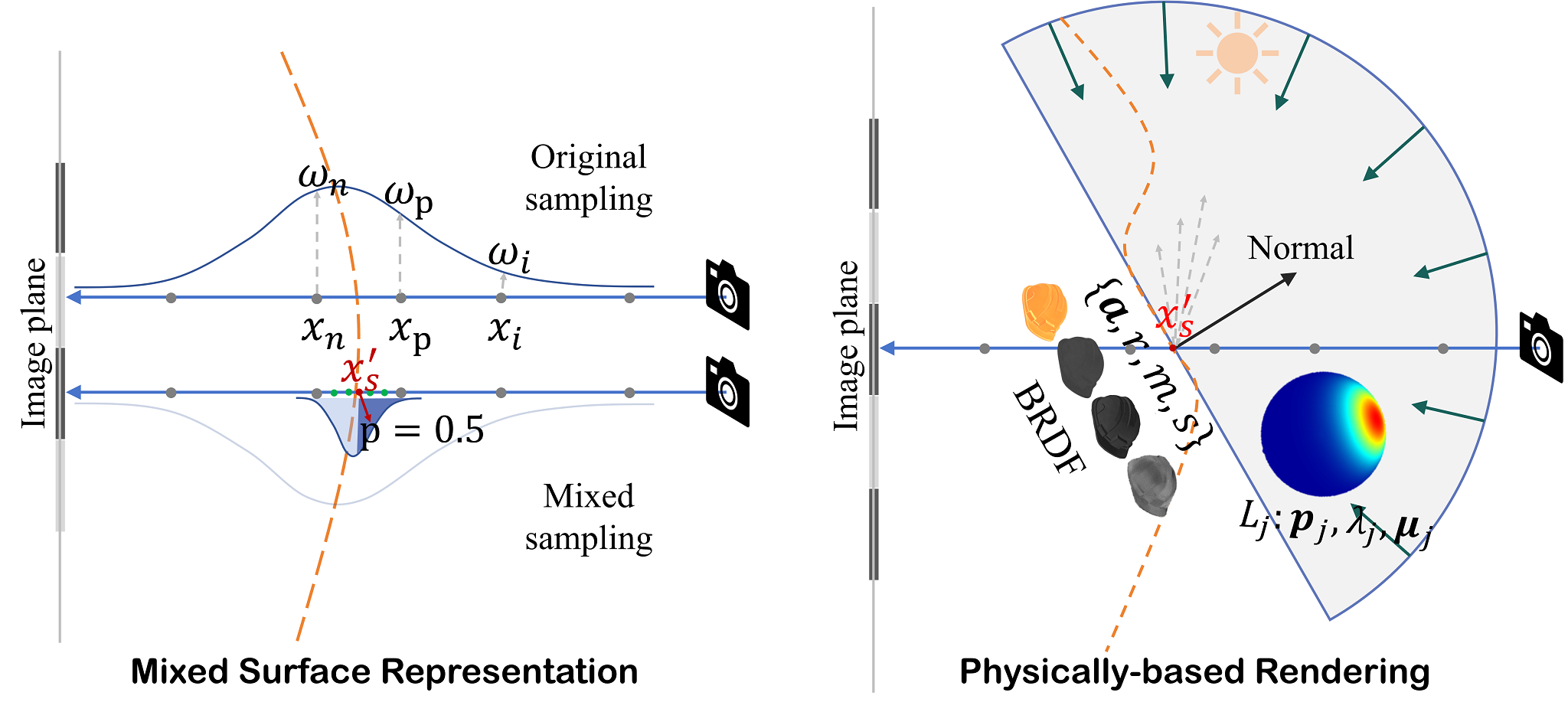}}
    \caption{Mixed surface representation and Physically-based Rendering implemented in our methd.}
    \vspace{-5mm}
	\label{fig:invrender}
\end{figure}

Empirically, we can represent an intersection point $\mathbf{x}_s$ lying on a ray $\mathbf{r}$ and iso-surface $\mathcal{S}$ simply, \ie, as a weighted sum of all the sampled candidates along this ray,
\begin{equation}\label{eq:iso_p0}
\vspace{-2mm}
    \mathbf{x}_s\triangleq \sum_{i=0}^N w_i\mathbf{x}_i, ~~\text{where }\mathbf{x}_i\triangleq \mathbf{r}_o + t_i\cdot \mathbf{r}_d.
\end{equation}
Note that, this formulation is based on the assumption guaranteed by Neus~\cite{neus2021} that the distribution of weights $\{w_i\}_i\in [0, 1]$ along a ray is well-approximated as a unimodal function whose value increases if $\mathbf{x}_i$ gets closer to $\mathcal{S}$ from the visible view (not back view). Unfortunately, we find that the resulted $\mathbf{x}_s$ is not smooth enough when it marches closed to $\mathcal{S}$, somehow suffering from the limitation of geometry consistency of generated sparse view images and materials in Sec.~\ref{sec:sprseview_gen}. Therefore, we introduce an simple yet efficient sampling method to extract more smooth $\{\mathbf{x}_s\}_s$ in an explicit way, as illustrated in Fig.~\ref{fig:invrender}.

\noindent
\textbf{Explicit Surface Sampling.}
With SDF $f_m(\theta)$ defined above, an intersection point $\mathbf{x}_s\in \mathbf{r}\cap\mathcal{S}$ lies between two specific sample points $\mathbf{x}_p, \mathbf{x}_n \in \mathbf{r}$, which are defined as:
\begin{align}
    \mathbf{x}_p\triangleq \mathop{\text{argmin}}\limits_{\mathbf{x}_i\in\mathbf{r}}f_m(\theta, \mathbf{x}_i), \forall~\mathbf{x}_i\in\mathbf{r}~s.t.~f(\theta, \mathbf{x}_i) \geq 0,\label{eq:x_p}\\
    \mathbf{x}_n\triangleq \mathop{\text{argmax}}\limits_{\mathbf{x}_i\in\mathbf{r}}f_m(\theta, \mathbf{x}_i), \forall~\mathbf{x}_i\in\mathbf{r}~s.t.~f(\theta, \mathbf{x}_i) \leq 0.\label{eq:x_n}
\end{align}
To ensure the visibility of $\mathbf{x}_s$, we choose $\mathbf{x}_p, \mathbf{x}_n$ with smallest depth $t_i$ following the z-buffer~\cite{z_buffer2001} method. 

With the selected two section points $\mathbf{x}_p, \mathbf{x}_n$, we uniformly sample $d$ (\eg, 8) points $\{\mathbf{x}_i'\}_i$ between them, and compute the weights of these samples by considering their SDF valuses and cosine similarity between gradients and view directions, except that we only interpolate an single point $\mathbf{x}_s'$ with probability $p=0.5$ w.r.t the cumulative distribution function (CDF) of $\{\mathbf{x}_i'\}_i$. Finally, the interpolated point $\mathbf{x}_s'$ can guarantee that $f_m(\theta, \mathbf{x}_s')\rightarrow 0$ during training with acceptable tolerance, and the smoothness is naturally enhanced.
since the intervals $\Delta\mathbf{x}_i'$ of samples $\{\mathbf{x}_i'\}_i$ is much less than the original intervals $\Delta\mathbf{x}_i$ along the ray.

This sampling method looks similar to the well-known ray marching~\cite{raymarching2021} that is also used in~\cite{physg2021, idr2020}. However, we have observed the slow convergence rate of ray marching if we directly apply it in our case. Moreover, the verbose iterations of ray-marching for searching the closet points to $\mathcal{S}$ is much more time-consuming than ours method.

\subsection{Physically-based Rendering}\label{sec:pbr_render}
\vspace{-3mm}
For training the multi-view generative model and the inverse rendering reconstruction, we both implement a physically-based rendering~\cite{pbr_3ed} approximation with the simplified Disney principle BRDF~\cite{pbr_disney2012} and spherical Gaussian~\cite{SGlight2009, SG_close2018} lighting as illustrated in Fig.~\ref{fig:invrender}.

\noindent
\textbf{The Rendering Equation.} To formulate the intersection of light and object's surface, the rendering equation~\cite{render_eq_1986} has been introduced based on the physical law of energy conservation, accounting the contribution of intrinsic materials of object. Given a surface point $\mathbf{x}$ (\ie, $\mathbf{x}_s$ above) with normal $\mathbf{n}$, the radiance of incident light $L_i$ shining at point $\mathbf{x}$ along incident direction $\omega_i$ is $L_i(\mathbf{\omega}_i, \mathbf{x})$, with an observation direction $\omega_o$ to the point $\mathbf{x}$, the reflection accounting for geometry and materials can be reduced by the bidirectional reflectance distribution function (BRDF) as $f_r({\omega}_o, {\omega}_i; \mathbf{x})$, the observed radiance $L_o({\omega}_o, \mathbf{x})$ is defined as,
\begin{equation}\label{eq:pbr_eq0}
L_o({\omega}_o, \mathbf{x}) = \int_{\Omega} {L_i({\omega}_i)} f_r({\omega}_o, {\omega}_i; \mathbf{x}) ({\omega}_i \cdot \mathbf{n}) d{\omega}_i,
\end{equation}
where $\Omega \triangleq \{{\omega}_i \,|\, {\omega}_i \cdot \mathbf{n} \geq 0\}$ is the hemisphere over which the integral Eqn.~\eqref{eq:pbr_eq0} is conducted.

\noindent
\textbf{Disney BRDF.}
The implementation of BRDF plays a core role of Eqn.~\eqref{eq:pbr_eq0}, we utilize the widely used Disney BRDF~\cite{pbr_disney2012} developed from Cook-Torrance~\cite{cook1981reflectance} as,
\begin{align}
    f_r({\omega}_o, {\omega}_i; \mathbf{x})&\triangleq f_d + f_s({\omega}_o, {\omega}_i; \mathbf{x})\\
    &=k_d\frac{\mathbf{a}}{\pi} + \frac{D(\mathbf{h}) F(\omega_o, \omega_i) G(\omega_o, \omega_i)}{4(\omega_o \cdot \mathbf{n})(\omega_i \cdot \mathbf{n})}
\end{align}
where $k_d$ is the diffuse refraction, $\mathbf{a}\in [0, 1]^3$ is the albedo, $\mathbf{h}=(\omega_o + \omega_i)/||\omega_o + \omega_i||_2$ is the half vector, $D, F, G$ are the normal distribution function (NDF), Fresnel and geometry term, respectively. The integral in Eqn.~\eqref{eq:pbr_eq0} can be approximately calculated either in discrete integration (\eg, precomputed radiance transfer (PRT))~\cite{prt_2002} or closed form~\cite{SGlight2009, SG_close2018}. To achieve more efficiency we adopt the implementation in closed form, which needs to be approximated as spherical Gaussians described as follows.

\noindent
\textbf{Spherical Gaussian Formulation.}
An spherical Gaussian (SG) in $\mathbb{R}^3$ is defined as~\cite{SGlight2009}:
\begin{equation}\label{eq:SG_def}
    G_s(\mathbf{v};\mathbf{p},\lambda,\bm{\mu)}\triangleq \bm{\mu} e^{\lambda(\mathbf{v}\cdot\mathbf{p}-1)},
\end{equation}
where $\mathbf{v}\in\mathbb{S}^2$ is the input vector, $\mathbf{p}\in \mathbb{S}^2$ is the lobe axis, $\lambda\in\mathbb{R}^+$ is the sharpness, and $\bm{\mu}\in \mathbb{R}_+^3$ is the amplitude.

\noindent
\textbf{Light as SG.} Concretely, the light $L_i$ is represented as a mixture of $N=16$ SGs as
\begin{equation}\label{eq:sg_li}
    L_i(\omega_i)\triangleq \sum_{j=1}^N G_s(\omega_i;\mathbf{p}_j,\lambda_j,\bm{\mu}_j).
\end{equation}

\noindent
\textbf{BRDF as SG.} The NDF term $D(\mathbf{h})$ can be approximately represented as a single spherically wrapped SG~\cite{SGlight2009},
\begin{equation}\label{eq:sg_wrap_D}
    D(\mathbf{h}) \approx G_s(\mathbf{h};\mathbf{p}^w, \lambda^w, \bm{\mu}^w),
\end{equation}
where $\mathbf{p}^w, \lambda^w, \bm{\mu}^w$ are wrapped as:
\begin{align}
    \mathbf{p}^w \triangleq2(\omega_o\cdot\mathbf{n})\mathbf{n}-\omega_o,
    \quad\lambda^w &\triangleq \frac{2}{r^4},
    \quad\bm{\mu}^w \triangleq \frac{1}{\pi r^4},
\end{align}
with $r\in [0, 1]$ is the roughness at point $\mathbf{x}$. Moreover, under the smooth and constant assumption~\cite{SGlight2009}, $F(\omega_o, \omega_i)$ and $G(\omega_o, \omega_i)$ can be calculated as constants $F_0, G_0$ on the support of $D(\mathbf{h})$ with approximation $\omega_i\approx 2(\omega_o\cdot\mathbf{n})\mathbf{n}-\omega_o$. Thus 
$f_s({\omega}_o, {\omega}_i; \mathbf{x})$ can be rewritten as
\begin{equation}\label{eq:sg_wrap_final}
    f_s({\omega}_o, {\omega}_i; \mathbf{x})\approx G_s(\mathbf{h};\mathbf{p}^w, \frac{\lambda^w}{4|\omega_o\cdot\mathbf{n}|}, F_0G_0\bm{\mu}^w).
\end{equation}
See more details about calculating $k_d, F_0, G_0$ w.r.t. $\omega_o, \omega_i, \mathbf{n}, r, m, s$ \footnote{$m, s$ are metallic and specular values at point $\mathbf{x}$, predicted by a materials MLP $f_c(\theta',\mathbf{x})=[\mathbf{a}, r, m, s]\in [0, 1]^6$.} in our supplemental materials.

\noindent
\textbf{Cosine as SG.} Now we consider the only left cosine $\omega_i\cdot \mathbf{n}$ in Eqn.~\eqref{eq:pbr_eq0}. As proposed in~\cite{SG_close2018}, it can be approximated as
\begin{equation}\label{eq:sg_cosine}
    \omega_i\cdot \mathbf{n}\approx G_s(\omega_i;\mathbf{n},0.0315,32.7080) - 31.7003.
\end{equation}

Finally, the integrand in Eqn.~\eqref{eq:pbr_eq0} is the product of three SGs, \ie Eqn.~\eqref{eq:sg_li},~\eqref{eq:sg_wrap_final} and~\eqref{eq:sg_cosine}, it's also a spherical Gaussian and can be integrated with closed form~\cite{SG_close2018}.


\subsection{Asset Enhancement}\label{sec:assts_3d}

\noindent
\textbf{Mesh Quadrification and UV Unwrapping.} The meshes generated by the Marching Cubes algorithm typically consist of millions of uneven triangles and messy topology, making it very difficult for artists to make any edits, such as reducing polygons to create additional levels of detail (LOD) variants. To overcome this challenge, we use Blender’s \cite{Blender} Quad Remesher \cite{quadriflow} tool to \textbf{\textit{remesh}} these triangular meshes into quad-faced meshes with a reasonable number of faces (e.g. $20k$), while preserving sharp edges and flat surfaces. Next, we \textbf{\textit{unwrap the UVs}} of the remeshed objects automatically, and \textbf{\textit{bake}} the per-vertex colors from the original high-poly model onto the remeshed low-poly model, ultimately converting it into a 3D asset that meets PBR standards, as shown in Fig.~\ref{fig:pipeline}. This whole process efficiently and reliably produces high-quality, refined 3D assets, facilitates the direct use of the generated digital assets within existing computer graphics (CG) workflows.
\section{Experiments}

\subsection{Implementation Details}
We wrote a Blender script to filter approximately $32,000$ 3D objects with complete PBR material maps from the Objaverse dataset \cite{deitke2023objaverse}, and then normalized all objects to a unit scale. To create a multi-view image dataset, cameras were placed in six positions: front, back, left, right, front-right, and front-left. We automated the modification of shader node connections in the $.glb$ format 3D objects to render multi-view images of color, normal, depth, albedo, roughness, and metallic.

In the multi-view synthesis stage, we fine-tuned on the pretrained Stable Diffusion Image Variants Model, which has image-to-image generation capabilities. We employed a batch size of $512$ and an image resolution of $256$, with the multi-view self-attention training for $30,000$ steps and the cross-domain attention training for $20,000$ steps. The entire training process was conducted on a single machine with $8$ A100 GPUs, taking nearly $5$ days.

In the inverse rendering phase, the SDF MLP $f_m(\theta)$ consists of 8 nonlinear layers of width 128, with a skip connection at
$4^{\text{th}}$ layer. The material MLP $f_c(\theta')$ comprises 4 nonlinear layers of width 128, concluding with Sigmoid activation layer. Positional encoding~\cite{nerf2020} is employed in both of $f_m(\theta)$ and $f_c(\theta')$, utilizing $L=10$ frequency components. For the parameters $\{\mathbf{p}_j,\lambda_j,\bm{\mu}_j\}_j$ of light $L_i$, we uniformly initialize $\{\mathbf{p}_j$ to be distributed on the unit sphere $\mathcal{S}^2$ and normalize $\bm{\mu}_j\}_j$ by dividing by the total energy.

\begin{figure}
	\centering
	{\includegraphics[width=1.0\linewidth]{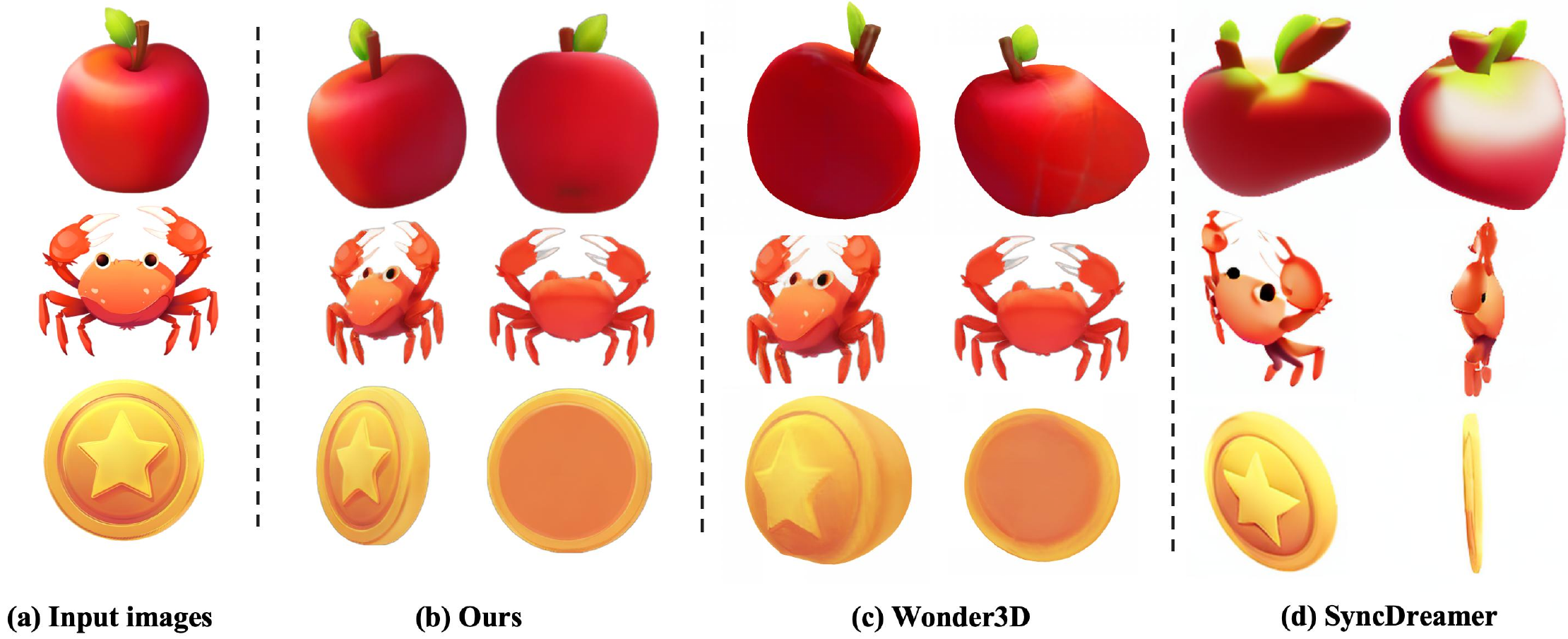}} 
    \captionof{figure}{The qualitative comparisons with baseline models on multi-view image generation.}
    \vspace{-3mm}
	\label{fig:nvs_compare}
\end{figure}

\subsection{Evaluation}
\noindent
\textbf{Baseline} We adopt Zero123 \cite{liu2023zero}, One-2-3-45++ \cite{liu2023one2345++}, SyncDreamer \cite{liu2023syncdreamer}, Wonder3D \cite{long2023wonder3d}, InstantMesh \cite{xu2024instantmesh}, SF3D \cite{sf3d2024}, and the latest work, 3DTopia-XL \cite{chen2024primx}, as baselines for image-to-3D generation. Zero123 \cite{liu2023one2345++} can generate novel views from any viewpoint as input. One-2-3-45++ quickly generates 3D content through a multi-view diffusion scheme. SyncDreamer \cite{liu2023syncdreamer} focuses on generating more consistent multi-view images. Wonder3D \cite{long2023wonder3d} extends the operational domain of the diffusion model to RGB and normal images. InstantMesh \cite{xu2024instantmesh}, SF3D \cite{sf3d2024}, and 3DTopia-XL \cite{chen2024primx} are categorized as 3D generative models.

\noindent
\textbf{Evaluation Datasets} As used in other works \cite{liu2023zero,liu2023syncdreamer}, we choose the Google Scanned Object (GSO) dataset \cite{downs2022google} for evaluation. It includes various everyday items, toys, and animals, and we have also added some plant and game prop images collected from the internet for visulization.

\noindent
\textbf{Metrics} For the first stage of novel-view synthesis, we use PSNR and SSIM \cite{wang2004image} to assess the quality of the generated color images. For the sparse view reconstruction task, We measure two metrics, Chamfer Distances (CD) and Volume IoU, between the reconstructed shape and ground truth.

\subsection{Novel View Synthesis}
We evaluated the quality of novel-view synthesis across different methods. Qualitative results can be seen in Fig.~\ref{fig:nvs} and Fig.~\ref{fig:nvs_compare}, while the quantitative outcomes are presented in Tab.~\ref{tab:nvs}. Wonder3D \cite{long2023wonder3d}, which lacks depth information as overall geometric prior, sometimes produces distorted geometries and struggles with complex structures. Despite SyncDreamer \cite{liu2023syncdreamer} introduced a volume attention scheme to enhance consistency, their model often yields unreasonable results. In contrast, our method faithfully generates 3D models according to the input images and performs well in terms of both geometry and texture.
\begin{table}[]
    \centering
    \scriptsize
    \resizebox{0.8\linewidth}{!}{
    \begin{tabular}{lccc}
       \toprule
       Method  & PSNR$\uparrow$ & SSIM$\uparrow$  \\
       \midrule
       Zero123 \cite{liu2023zero}    
       & $18.64$ & $0.796$    \\
       SyncDreamer \cite{liu2023syncdreamer}   
       & $20.05$ & $0.803$  \\
       Wonder3D \cite{long2023wonder3d}
       &  $24.11$ & $0.893$    \\
       Ours    
       & \textbf{27.93} & \textbf{0.937} \\
       \bottomrule
    \end{tabular}
    }
    \vspace{-2mm}
    \caption{The quantitative comparison on novel view synthesis.}
    \vspace{-1mm}
    \label{tab:nvs}
\end{table}

\subsection{Surface Reconstruction}
\begin{table}[]
    \centering
    \scriptsize
    \resizebox{0.8\linewidth}{!}{
    \begin{tabular}{lccc}
       \toprule
       Method & Chamfer Dist. $\downarrow$ & Volume IoU $\uparrow$ \\
       \midrule
       Zero123 \cite{liu2023zero}   
       & $0.0342$ & $0.5033$  \\
       One-2-3-45++ \cite{liu2023one2345++}    
       & $0.0274$ & $0.5433$ \\
       SyncDreamer~\cite{liu2023syncdreamer}   
       & $0.0249$ & $0.5301$  \\
       Wonder3D \cite{long2023wonder3d}
       & $0.0237$ & $0.5762$  \\
       InstantMesh \cite{xu2024instantmesh}
       & $0.0246$ & $0.5591$  \\
       SF3D \cite{sf3d2024}
       & $0.0311$ & $0.5203$  \\
       3DTopia-XL \cite{chen2024primx}
       & $0.0378$ & $0.5126$  \\
       Ours
       &  $\mathbf{0.0231}$  & $\mathbf{0.5779}$ \\
       \bottomrule
    \end{tabular}}
    \vspace{-2mm}
    \caption{Quantitative comparison on single view reconstruction.}
    \vspace{-2mm}
    \label{tab:recon}
\end{table}
We evaluate the effectiveness of the mixed representation of surface proposed in Sec.~\ref{sec:explit_iso}, which enhances the smoothness and contributes to a more visually appealing surface. Compared to other methods, our approach achieves a SOTA surface accuracy, as shown in Tab.~\ref{tab:recon} and Fig.~\ref{fig:recon}.


\begin{figure}
	\centering
	{\includegraphics[width=1.0\linewidth]{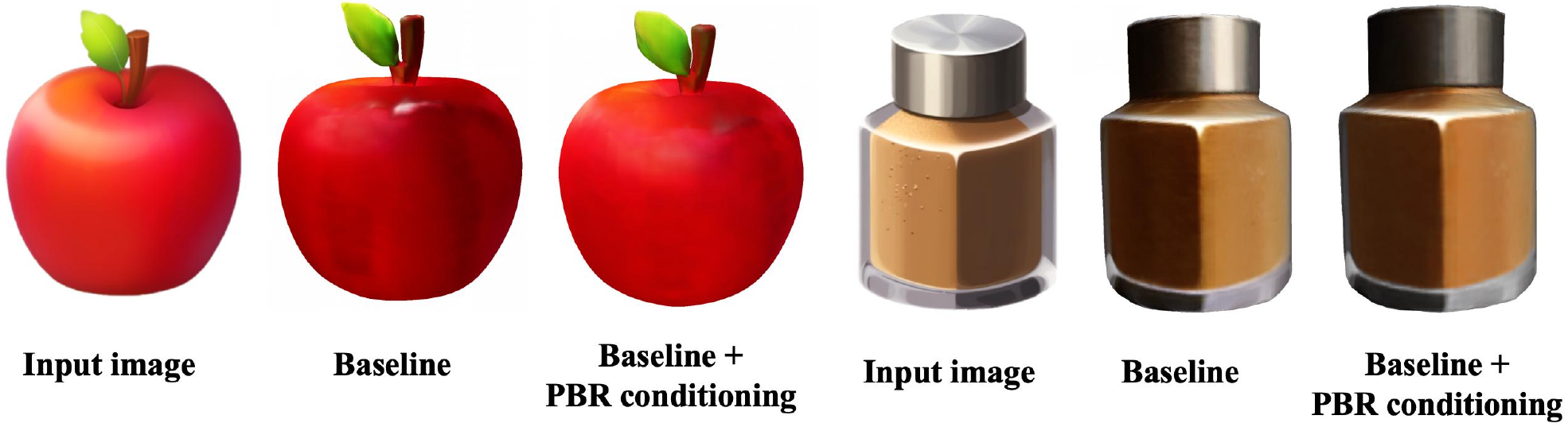}}
    \vspace{-3mm}
    \captionof{figure}{By ensuring the correct application of PBR conditions in both the multi-view image synthesis stage and the inverse rendering stage, our model efficiently distinguishes shadows and highlights on the input images (as seen in the \textbf{\textit{apple}} and \textbf{\textit{bottle cap}}). It also improves the reconstruction results for transparent materials (\textbf{\textit{bottle body}}) to some extent.}
    \vspace{-5mm}
	\label{fig:ablation}
\end{figure}

\subsection{Materials and Relighting}



\begin{figure}
	\centering
    \begin{subfigure}{0.22\linewidth}
        \includegraphics[width=\linewidth]{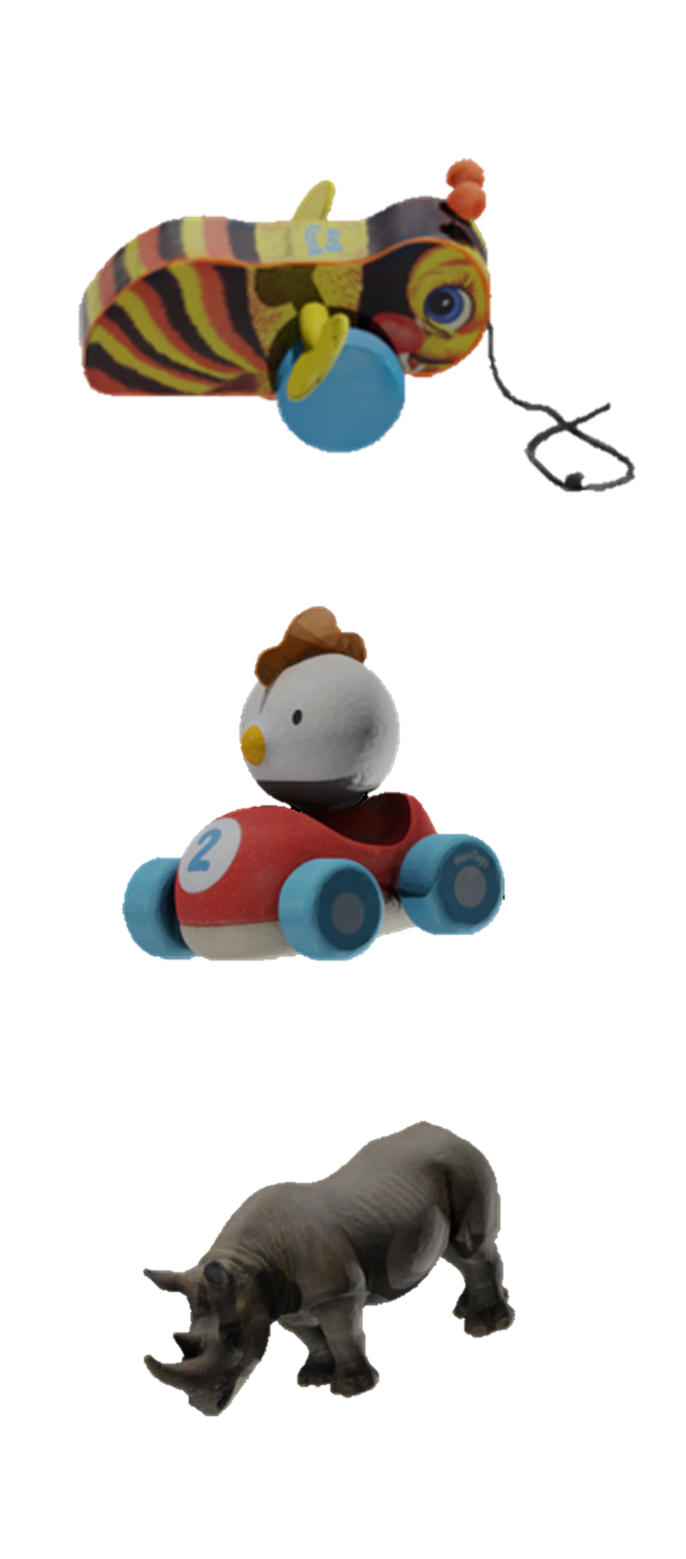}
        \caption{Input image}
    \end{subfigure}
    \begin{subfigure}{0.23\linewidth}
        \includegraphics[width=\linewidth]{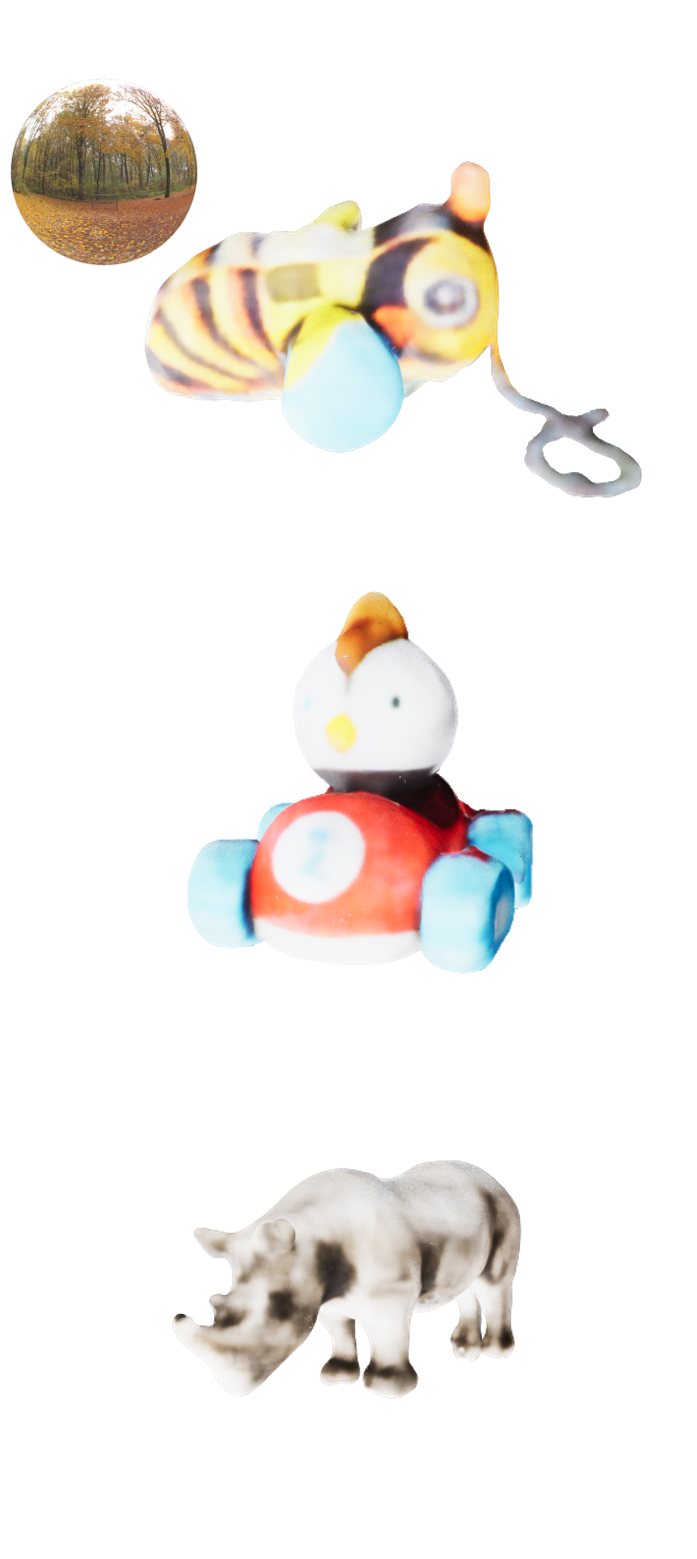}
        \caption{Bright light}
    \end{subfigure}
    \begin{subfigure}{0.23\linewidth}
        \includegraphics[width=\linewidth]{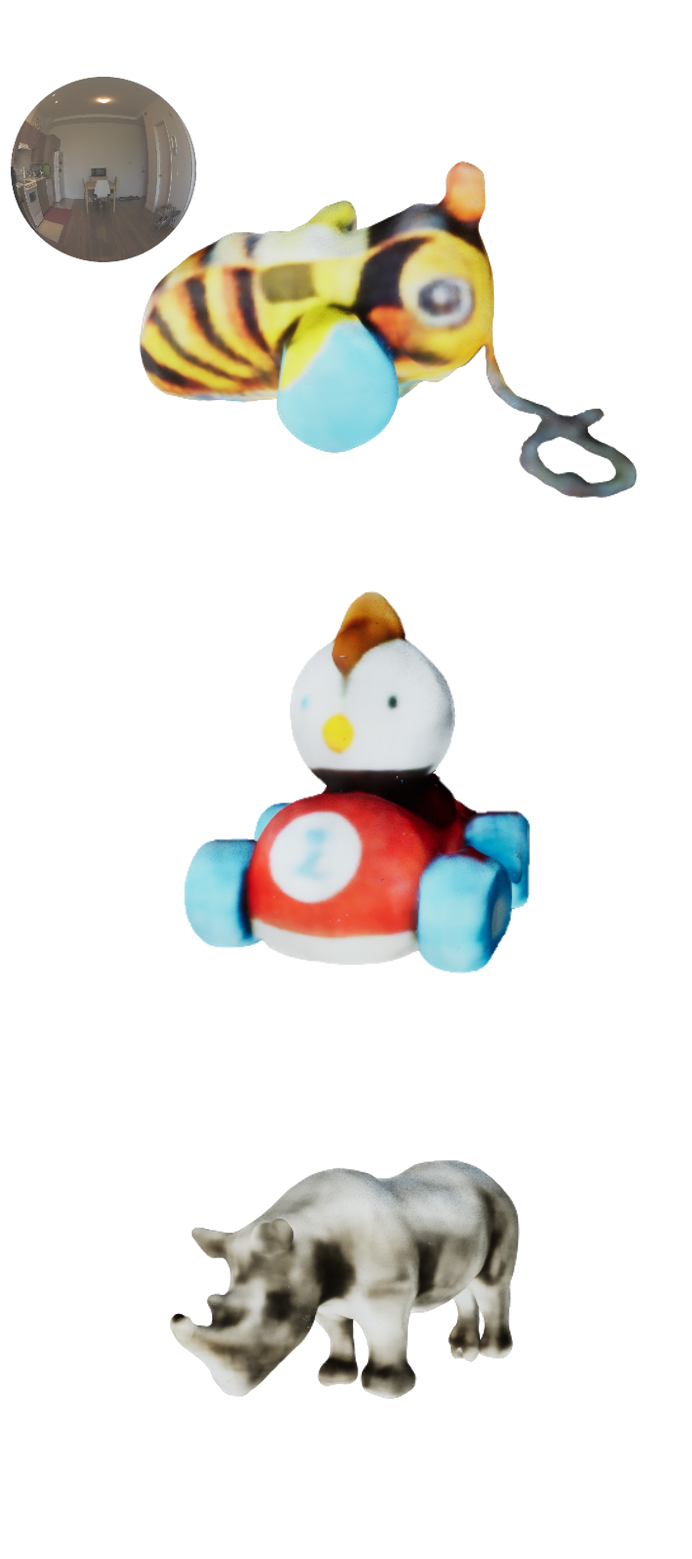}
        \caption{Midtone light}
    \end{subfigure}
    \begin{subfigure}{0.23\linewidth}
        \includegraphics[width=\linewidth]{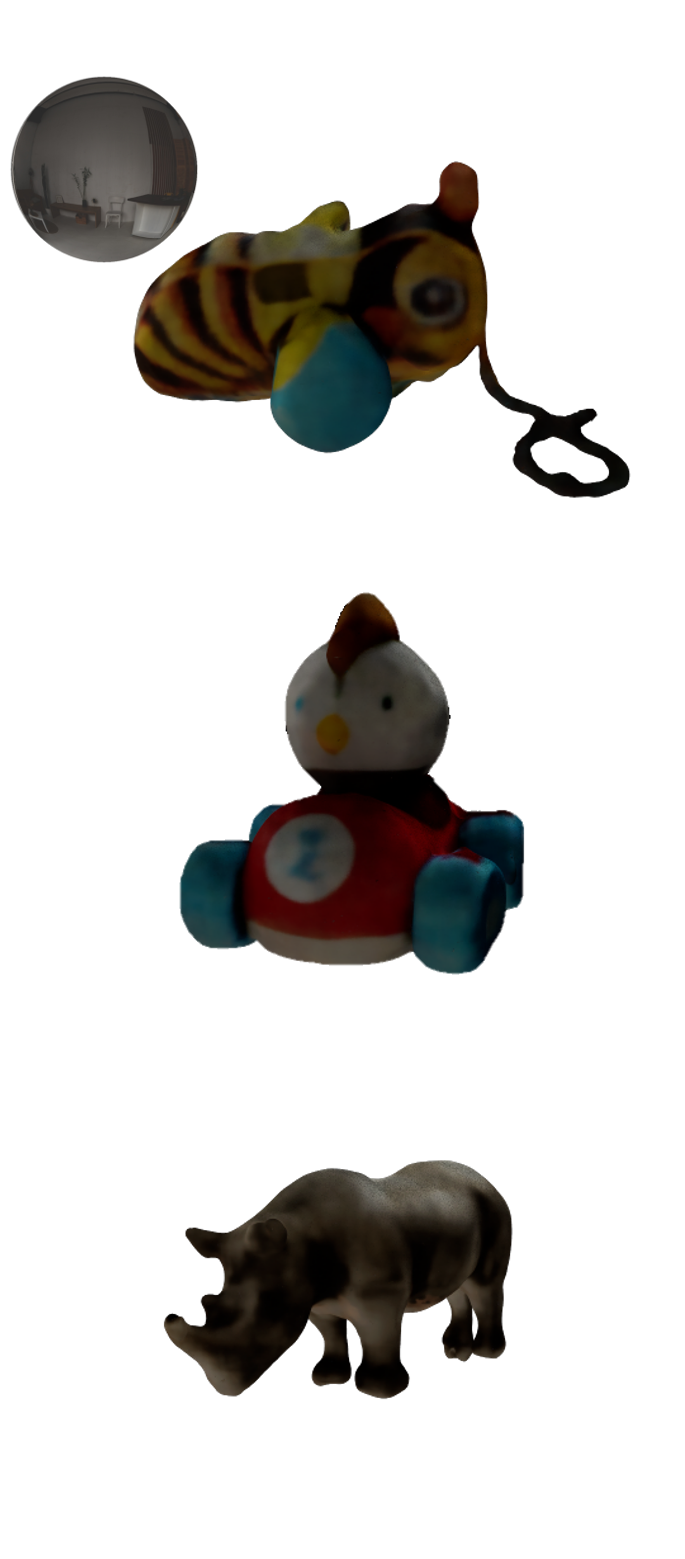}
        \caption{Dark light}
    \end{subfigure}
	\vspace{-3mm}
	\caption{Relighting of the 3D objects w.r.t. varying lighting environment maps from bright to dark.}
    \vspace{-5mm}
	\label{fig:relighting}
\end{figure}

We can naturally relight the 3D objects using the materials integrated into the complete 3D surface mesh following the Physically Based Rendering (PBR) process, which has been mapped into UV space. We employ three types of environment maps to simulate dark, ordinary, and bright conditions, respectively. In Fig.~\ref{fig:ablation}, we demonstrate that our model can handle special cases such as highlights and metallic materials. And as shown in Fig.~\ref{fig:relighting}, it evaluates how the intrinsic materials extracted by our method interact with varying light environment maps.

\section{Conclusions}
In this paper, we introduce GraphicsDreamer, an advanced workflow for 3D objects modeling tailored for modern graphics. It comprises a multi view diffusion generative model that integrates geometry and intrinsic materials, alongside a deep learning-based inverse rendering that aggregates multi view images and intrinsic materials into a complete 3D surface mesh.
By employing Physically Based Rendering (PBR) as a condition in both stages, GraphicsDreamer ensures convincing consistency in materials throughout the entire process.
Furthermore, GraphicsDreamer incorporates topology optimization and UV unwrapping, which are often overlooked in previous works. Experiments demonstrate that our method excels in fidelity and accuracy, particularly regarding geometric and textural details, while also ensuring clean topology. 


In future work, we plan to increase the amount of training data and simultaneously render objects with a wider variety of lighting conditions using thousands of environment maps. With the improved re-training of generative intrinsic materials, we aim to achieve a more stable and accurate inverse rendering refinement process.
{
    \small
    \bibliographystyle{ieeenat_fullname}
    \bibliography{main}
}

\clearpage
\clearpage
\setcounter{page}{1}
\maketitlesupplementary

\setcounter{table}{0}
\setcounter{section}{0}
\renewcommand\thesection{A\arabic{section}}

In this supplementary document, we first introduce the creation of the training dataset used for generating PBR materials. Then, we outline some implementation details of inverse rendering, and finally, we present more visual results.

\section{Dataset Preparation}
\begin{figure*}
	\centering
	{\includegraphics[width=0.7\linewidth]{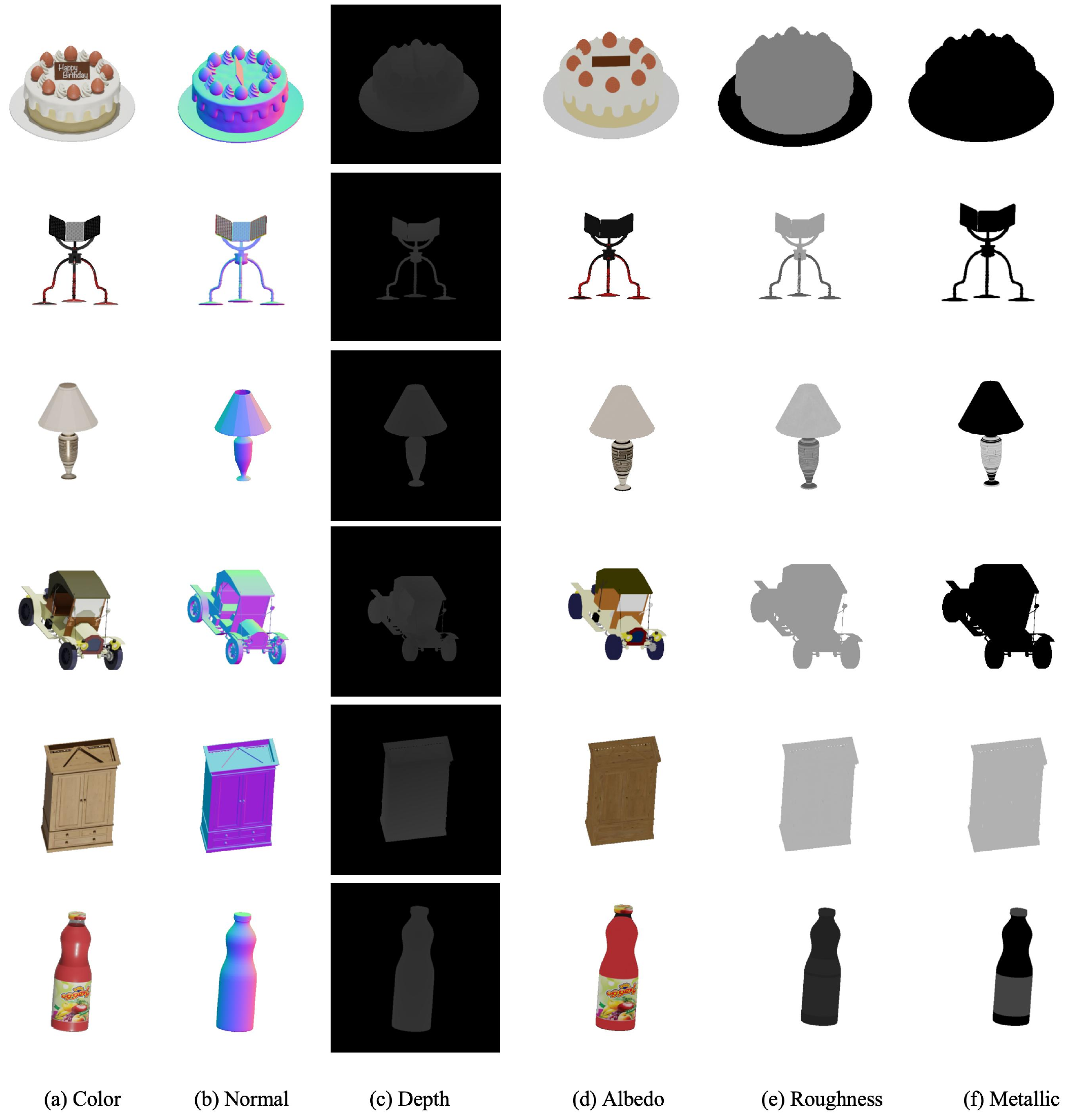}} 
    \captionof{figure}{Multi-view PBR material images, created by rendering the Objaverse dataset. \cite{deitke2023objaverse}.}
    \vspace{-2mm}
	\label{fig:dataset_pics}
\end{figure*}

\begin{figure*}[tp!]
	\centering
	{\includegraphics[width=0.9\linewidth]{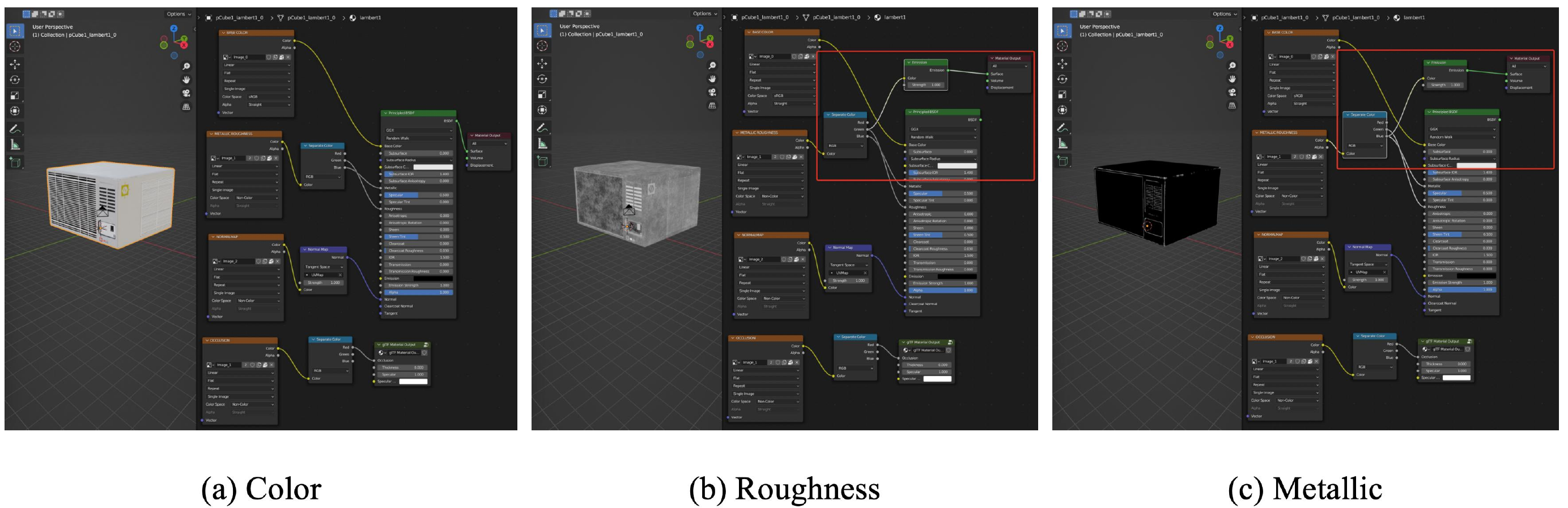}} 
    \captionof{figure}{Using a Blender script, we automatically manipulate shader nodes of 3D objects to obtain the corresponding PBR material images, which are then output through the added \textit{ShaderNodeEmission} node.}
    \vspace{-5mm}
	\label{fig:dataset_blender}
\end{figure*}

To make full use of the material information embedded within the \textit{.glb} format 3D assets from the Objaverse dataset \cite{deitke2023objaverse}, and to supply our cross-domain PBR material generation model with high-quality training data, we wrote a Blender \cite{Blender} script and filtered out approximately $32,000$ 3D objects with complete PBR materials. After normalizing them to a unit size, we rendered these objects from six viewpoints: front, back, left, right, front-right, and front-left, to obtain multi-view images of color, normal, depth, albedo, roughness, and metallic, as shown in Fig.~\ref{fig:dataset_pics}. Specifically, the first four types of images - color, normal, depth, and albedo, can be directly output from its existing shader nodes, while roughness and metallic images are obtained by automatically modifying the connections of shader nodes and output from an added \textit{ShaderNodeEmission} node, see Fig.~\ref{fig:dataset_blender}. We will release the dataset creation code upon acceptance of this paper.

\section{Details of Inverse Rendering}
\label{sec:detail_pbr}
In this section, we demonstrate the implementation details of inverse rendering,  which includes sampling the explicit surface in Sec.~(3.2) and calculating the simplified Disney BRDF in Sec.~(3.3.).

\subsection{Explicit Surface Sampling}

{\bf{Z-buffer Implementation.}}
As noted in lines $335-367$, for each ray $\mathbf{r}$, we will first find two specific sample points $\mathbf{x}_p,~\mathbf{x}_n\in\mathbf{r}$ following the z-buffer~\cite{z_buffer2001} method. Since the original z-buffer method is used for rasterizing a mesh surface, we have implemented it with specific revisions in our work.

Concretely, given $N$ (\eg, $N=64$) sample points $\{\mathbf{x}_i\}_{i=1}^N$ along the ray $\mathbf{r}$, with the definition of $\mathbf{x}_i$ in Eqn.~(4) in line $343$, we reorder $\{\mathbf{x}_i\}_i$ w.r.t. an increasing order of $\{t_i\}_i$. Next, we compute the SDF values of $\{\mathbf{x}_i\}_i$ with the SDF MLPs as $\{f_m(\theta,\mathbf{x}_i)\}_i$, which hold positive values in the exterior space and negative values in the interior space. We also compute the signs of these SDF values as $\{\text{sign}(f_m(\theta,\mathbf{x}_i))\}_i$. Therefore, we have

\begin{equation}\label{eq:indicator}
 \text{sign}(f_m(\theta,\mathbf{x}_i))=\left\{  
 \begin{array}{lr}
 \hspace{-1mm} +1,~~\mathbf{x}_i~{\text{in exterior space}}, &  \\  
 \hspace{-1mm} -1,~~\mathbf{x}_i~{\text{in interior space}}.\\    
 \end{array}  
 \right.
 \end{equation}
 
 By this way, the product of signs of two neighbor points $\mathbf{x}_{i-1},~\mathbf{x}_i$ as $\text{sign}(f_m(\theta,\mathbf{x}_{i-1}))\cdot \text{sign}(f_m(\theta,\mathbf{x}_i))$ equals $-1$ means that the there exists an intersection point $\mathbf{x}_s$ between $\mathbf{x}_{i-1}$ and $\mathbf{x}_i$. The rule of z-buffer requires that the intersection point $\mathbf{x}_s$ has the minimal depth $t_s$ along the ray $\mathbf{r}$. Fortunately, since the depth values $\{t_i\}_i$ are arranged in increasing order w.r.t the indices $i=0,1,...N-1$, we can compute the desired pair $(\mathbf{x}_{i-1}, \mathbf{x}_i)$ at the first instance where $\text{sign}(f_m(\theta,\mathbf{x}_{i-1}))\cdot \text{sign}(f_m(\theta,\mathbf{x}_i)) == -1$ occurred. This can be efficiently implemented using the $\text{argmax}()$ or $\text{argmin}()$ operator.
Thus, we have indeed identified a pair $(\mathbf{x}_p, \mathbf{x}_n)\triangleq (\mathbf{x}_{i-1}, \mathbf{x}_i)$.

\subsection{Disney BRDF Calculation}
As demonstrated in lines $421-433$, we calculate the simplified Disney BRDF as follows.

Given an intersection point $\mathbf{x}_s$ sampled above, we derive the normal of $\mathbf{x}_s$ as the normalized gradient of SDF MLPs $f_m(\theta,\cdot)$ as
\begin{equation}
    \mathbf{n}=\frac{\nabla_{\mathbf{x}} f_m(\theta, \mathbf{x}_s)}{||\nabla_{\mathbf{x}} f_m(\theta, \mathbf{x}_s)||_2}.
\end{equation}
Also, we utilize the materials MLPs $f_c(\theta', \cdot)$ to compute the correspondent intrinsic materials as $f_c(\theta', \mathbf{x}_s)=[\mathbf{a}, r, m, s] \in [0, 1]^6$.
Next, we use the following expressions to calculate the terms of BRDF inspired by prior work~\cite{pbr_disney2012}.

For the diffuse refraction $k_d$ in Eqn.~(9) in line $403$, we compute
$k_d=(1-m)k_d^ik_d^o$ with
\begin{align}
    \omega_i &\approx 2(\omega_o\cdot\mathbf{n})\mathbf{n}-\omega_o,\\
    \mathbf{h} &= \frac{\omega_i + \omega_o}{||\omega_i + \omega_o||_2},\\
    F_{D90} &= 0.5 + 2 (\omega_i\cdot\mathbf{h})^2 r,\\
    k_d^i &= 1 + (F_{D90} - 1)(1 - \omega_i\cdot \mathbf{n})^5,\\
    k_d^o &= 1 + (F_{D90} - 1)(1 - \omega_o\cdot \mathbf{n})^5.
\end{align}

For the Fresnel term $F_0$ in Eqn.~(14) in line $431$, we compute it by
\begin{align}
    C_s &= (1-m)s + m\mathbf{a},\\
    F_0 &= C_s + (1-C_s)(1- \omega_i\cdot\mathbf{h})^5.
\end{align}

For the geometry term $G_0$ in Eqn.~(14) in line $431$, we compute it by
\begin{align}
    k &= \frac{(r+1)^2}{8},\\
    G_0 &= \frac{\omega_i\cdot \mathbf{n}}{\omega_i\cdot \mathbf{n}(1-k)+k}\cdot\frac{\omega_o\cdot \mathbf{n}}{\omega_o\cdot \mathbf{n}(1-k)+k}.
\end{align}



\section{More Results}
\begin{figure*}[tp!]
	\centering
	{\includegraphics[width=\linewidth]{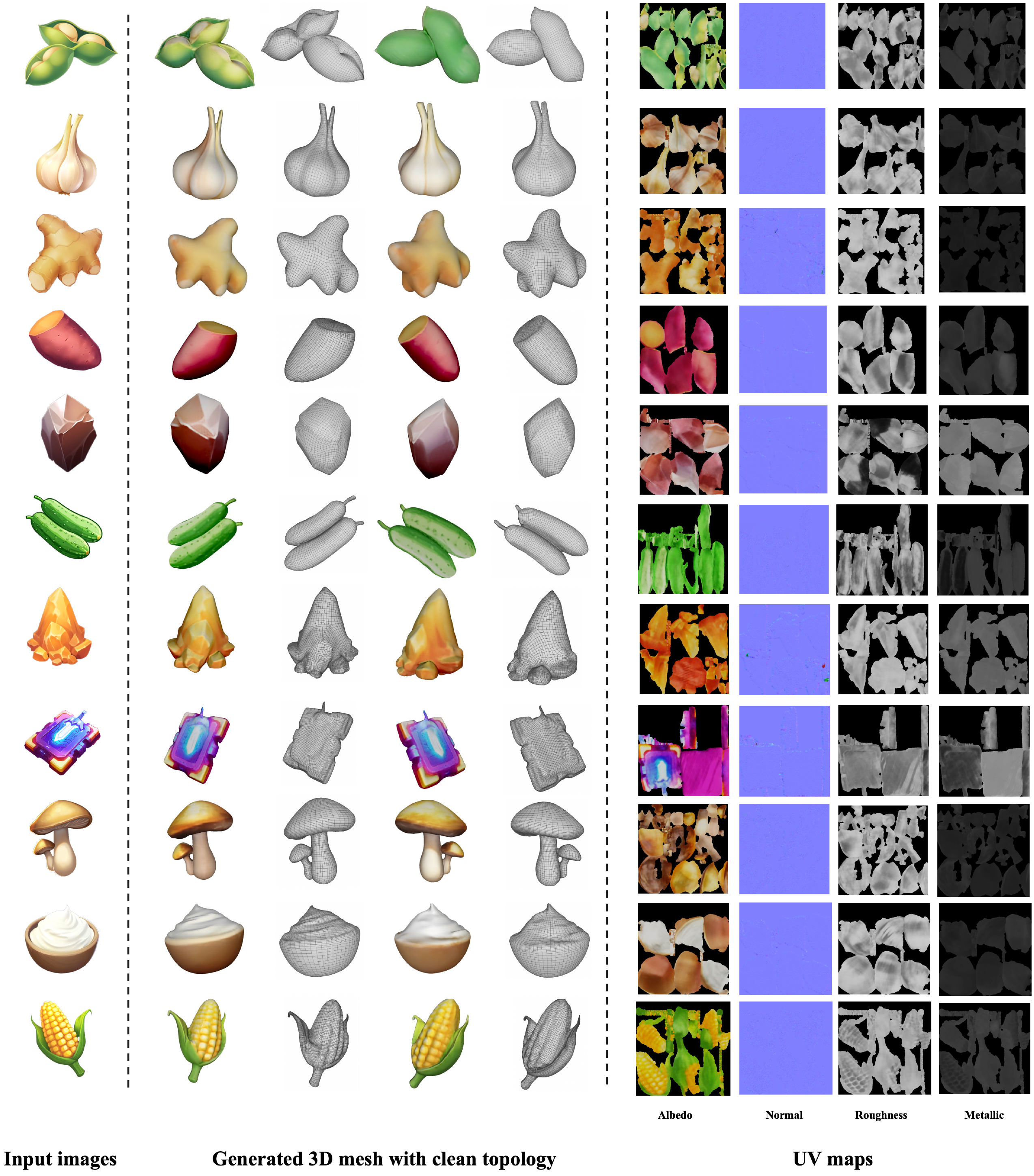}} 
    \captionof{figure}{More visual results on various game prop images collected from the Internet.}
	\label{fig:more_results1}
\end{figure*}

\begin{figure*}[tp!]
	\centering
	{\includegraphics[width=\linewidth]{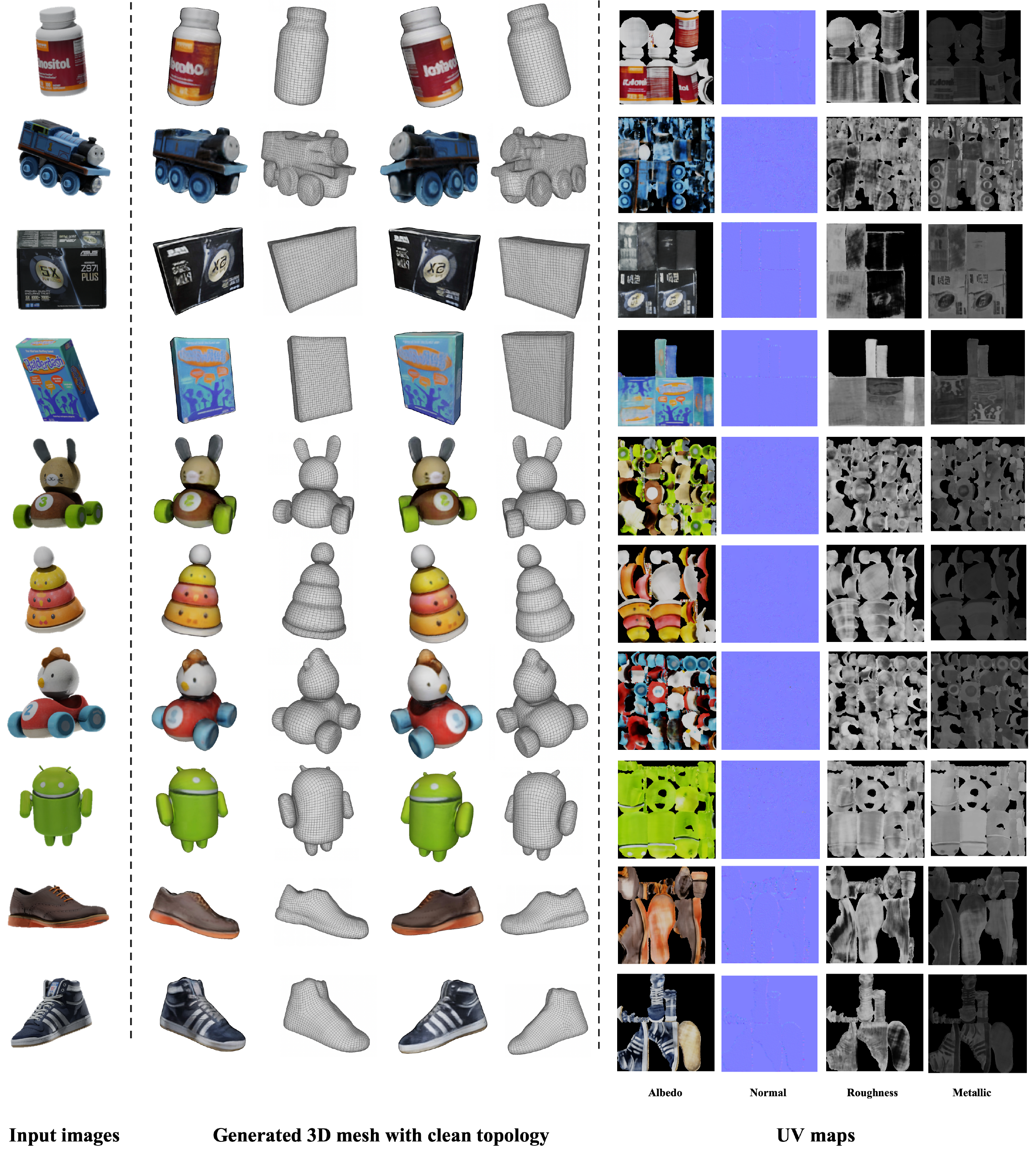}} 
    \captionof{figure}{More visual results on GSO dataset \cite{downs2022google}.}
	\label{fig:more_results2}
\end{figure*}

We enhance the generated 3D objects by automated remeshing, UV unwrapping, and baking, producing 3D assets that can be directly imported into graphics engines. More results can be seen in Fig.~\ref{fig:more_results1} and Fig.~\ref{fig:more_results2}.

\end{document}